\documentclass[12pt]{article}

\def\ltwid{\mathrel{\raise.3ex\hbox{$<$\kern-.75em\lower1ex\hbox{$\sim$}}}}
\def\gtwid{\mathrel{\raise.3ex\hbox{$>$\kern-.75em\lower1ex\hbox{$\sim$}}}}
\def\comp{{\rm C}\llap{\vrule height7.1pt width1pt depth-.4pt\phantom t}}

\def\square{\kern1pt\vbox{\hrule height 1.2pt\hbox{\vrule width 1.2pt\hskip 3pt
   \vbox{\vskip 6pt}\hskip 3pt\vrule width 0.6pt}\hrule height 0.6pt}\kern1pt}

\begin{document}

\begin{titlepage}

\begin{flushright}
ITP-UU-11/29, SPIN-11/22 \\ CCTP-2011-22, UFIFT-QG-11-06
\end{flushright}

\vskip .4cm

\begin{center}
{\bf Gauging away Physics}
\end{center}

\begin{center}
S. P. Miao$^*$
\end{center}

\begin{center}
\it{Institute for Theoretical Physics \& Spinoza Institute, Utrecht
University \\ Leuvenlaan 4, Postbus 80.195, 3508 TD Utrecht, NETHERLANDS}\\
\end{center}

\begin{center}
N. C. Tsamis$^{\dagger}$
\end{center}

\begin{center}
\it{Institute of Theoretical Physics \& Computational Physics, and \\
Department of Physics, University of Crete \\
 GR-710 03 Heraklion, HELLAS}
\end{center}

\begin{center}
R. P. Woodard$^{\ddagger}$
\end{center}

\begin{center}
\it{Department of Physics, University of Florida \\
Gainesville, FL 32611, UNITED STATES}
\end{center}

\begin{center}
ABSTRACT
\end{center}
We consider the recent argument by Higuchi, Marolf and Morrison
\cite{Don} that a nonlocal gauge transformation can be used to
eliminate the infrared divergence of the graviton propagator, when
evaluated in Bunch-Davies vacuum on the open coordinate submanifold
of de Sitter space in transverse-traceless-synchronous gauge.
Because the transformation is not local, the equal time commutator of
undifferentiated fields no longer vanishes. From explicit
examination of the Wightman function we demonstrate that the
transformation adds anti-sources in the far future which cancel the
bad infrared behavior but also change the propagator equation. The
same problem exists in the localized version of the recent argument.
Adding such anti-sources does not seem to be legitimate and could be
used to eliminate the infrared divergence of the massless, minimally
coupled scalar. The addition of such anti-sources in flat space QED
could be effected by an almost identical gauge transformation, and
would seem to eliminate the well known infrared divergences which
occur in loop corrections to exclusive amplitudes.

\begin{flushleft}
\end{flushleft}

\begin{flushleft}
$^*$ e-mail: S.Miao@uu.nl \\
$^{\dagger}$ e-mail: tsamis@physics.uoc.gr \\
$^{\ddagger}$ e-mail: woodard@phys.ufl.edu
\end{flushleft}

\end{titlepage}

\section{Introduction}

There is a wide disparity of opinion concerning quantum field theory
in the de Sitter geometry between those who approach the subject
from the perspective of cosmology and those who view it from the
perspective of mathematical physics. Cosmologists tend to see the de
Sitter geometry as an idealized special case of the larger class of
spatially flat, Friedman-Robertson-Walker (FRW) geometries, whose
invariant element in conformal coordinates is,
\begin{equation}
ds^2_{\rm FRW} = a^2(\eta) \Bigl[ -d\eta^2 + d\vec{x} \!\cdot\!
d\vec{x}\Bigr] \qquad {\rm de\ Sitter} \Longrightarrow a(\eta) =
-\frac{\ell}{\eta} \; . \label{invelm}
\end{equation}
Many mathematical physicists find this view antipathetic.
To them de Sitter has a privileged status as the unique, maximally
symmetric solution to the Einstein equations with positive
cosmological constant ($\Lambda = (d-1)/\ell^2$ in $d$ spacetime
dimensions), and the explication of its properties requires the use
of coordinates which cover the full de Sitter manifold.

The matter of coordinate symmetries illustrates the contrasting
perspectives. The key symmetries for cosmologists are
homogeneity and isotropy,
\begin{equation}
\matrix{ \eta' = \eta \cr {x'}^i = x^i + \epsilon^i} \qquad , \qquad
\matrix{\eta' = \eta \cr {x'}^i = R^{ij} x^j } \; . \label{HI}
\end{equation}
Mathematical physicists attach equal importance to dilatations and
spatial special conformal transformations,
\begin{equation}
\matrix{\eta' = k \eta \cr {x'}^i = k x^i} \qquad , \qquad
\matrix{\eta' = \frac{\eta}{1 - 2 \vec{\theta} \cdot \vec{x} +
\theta^2 x \cdot x} \cr {x'}^i = \frac{x^i - \theta^i x \cdot x}{1 -
2 \vec{\theta} \cdot \vec{x} + \theta^2 x \cdot x} } \; , \label{DS}
\end{equation}
where $x \cdot x \equiv -\eta^2 + \vec{x} \cdot \vec{x}$ and
$\theta^2 \equiv \vec{\theta} \cdot \vec{\theta}$. The union of
(\ref{HI}) and (\ref{DS}) comprises the full de Sitter group, and
mathematical physicists typically expect that it should play the
same role in organizing quantum field theory on de Sitter space as
the Poincar\'e group does for Minkowski space. Cosmologists are not
especially concerned with (\ref{DS}) because their interest in de
Sitter is as an approximation to primordial inflation, during which
the slight time dependence of the Hubble parameter breaks these
symmetries.

For most quantum fields there is not much practical difference
between the two viewpoints. Fields with positive mass-squared, and
massless, conformally invariant fields admit a de Sitter invariant
formulation. However, even as it is for flat space, one gets the
correct answer by following canonical quantization, without taking
advantage of coordinate symmetries. The fields which occasion
conflict are those which are both massless and not conformally
invariant: massless, minimally coupled (MMC) scalars and the
graviton. These are the fields responsible for primordial
perturbations in the theory of inflation \cite{Pperts}.
In the de Sitter limit, the unique de Sitter invariant wave function
for these fields is Bunch-Davies vacuum \cite{BD}. The scalar and
tensor power spectra are defined in this vacuum, and it is obvious
from their scale invariance (which becomes exact in the de Sitter
limit) that the Fourier mode sums for the propagators of both fields
must diverge in the infrared \cite{KOW}. What to make of this has
provoked decades of controversy.

For MMC scalars there is general agreement that the infrared
divergence means the Bunch-Davies wave function is not
normalizable and hence not a state. The de Sitter problem can be
viewed as but a special case of the infrared divergences discovered
in 1977 by Ford and Parker \cite{FP} for cosmologies with deceleration
parameter less than or equal to the matter-dominated value of $q =
+\frac12$. (de Sitter has $q = -1$, and data constrains single-scalar
inflation to the range $-1 < q < -0.986$ \cite{KOW}.) Infrared
divergences signal
something unphysical about the question being posed. For example, the
infrared divergences in loop corrections to the exclusive amplitudes
of flat space quantum electrodynamics (QED) arise because no real
experiment is able to exclude the possibility of a small amount of
energy escaping detection in the form of very soft photons. When one
computes inclusive rates and cross sections the results are infrared
finite but they display a measured dependence upon the detector
sensitivity \cite{BNSW}. For cosmologies with $q \leq +\frac12$ the
problem is that one cannot measure the super-horizon modes to be in
perfect Bunch-Davies vacuum, or anything else. Once the assumption
of Bunch-Davies vacuum is relaxed for the super-horizon modes
\cite{fixes} one gets finite propagators which, however, show
secular growth \cite{JMPW}.

It will be noted that the infrared problem of MMC scalars in de Sitter
rests comfortably within the larger context of spatially flat, FRW
geometries, just as cosmologists prefer. That does not mean there
has been no special attention to de Sitter. A famous result was
derived in 1982 for the secular --- and de Sitter breaking ---
growth of the coincidence limit of the MMC scalar propagator by Vilenkin
and Ford, by Linde and by Starobinsky \cite{VFLS}. Mathematical physicists
were finally persuaded of the reality of de Sitter breaking for this
system by the formal proof Allen and Folacci presented in 1987
\cite{AF}. However, there has been a notable tendency among the
mathematically inclined to discount de Sitter breaking because the
expectation value of the free stress tensor is de Sitter invariant.
That is the result of derivatives in the free stress tensor, and
merely postpones the appearance of secular, de Sitter breaking
effects to loop corrections in any theory which possesses
nonderivative interactions \cite{phi4,Yukawa,SQED}.

No similar consensus has been reached for gravitons, which we define
as the perturbation of the full metric from its background value,
\begin{equation}
h_{\mu\nu}(\eta,\vec{x}) \equiv \frac1{\sqrt{32\pi G}} \Bigl[
g_{\mu\nu}(\eta,\vec{x}) - a^2(\eta) \eta_{\mu\nu} \Bigr] \equiv
a^2(\eta) \widetilde{h}_{\mu\nu}(\eta,\vec{x}) \; .
\end{equation}
Because we shall have no more need for the full metric, we
henceforth employ the symbol $g_{\mu\nu}(\eta,\vec{x})$ to describe
the background metric, $g_{\mu\nu}(\eta,\vec{x}) \equiv a^2(\eta)
\eta_{\mu\nu}$. Gravitons are closely related to MMC scalars
because Grishchuk \cite{Grishchuk} long ago showed that conformally
rescaled graviton fields which are purely spatial, transverse and
traceless,
\begin{equation}
\widetilde{h}_{\mu 0}(\eta,\vec{x}) = 0 \quad , \quad \partial_i
\widetilde{h}_{ij}(\eta,\vec{x}) = 0 \quad , \quad
\widetilde{h}_{ii}(\eta,\vec{x}) = 0 \; , \label{TTS}
\end{equation}
obey exactly the same linearized equation of motion as the MMS
scalar,
\begin{equation}
\partial_{\mu} \Bigl[ \sqrt{-g} g^{\mu\nu} \partial_{\nu}
h_{ij}(\eta,\vec{x}) \Bigr] \equiv \sqrt{-g} \, \square
\widetilde{h}_{ij}(\eta,\vec{x}) = 0 \; . \label{EOM}
\end{equation}
(Note that we really do mean $\square$ to be the covariant scalar
d'Alembertian, even though it acts on a tensor field.) Just as in
flat space, one can prove that there are no other dynamical gravitons
by first solving the full linearized field equations in some volume
gauge condition --- for example, de Donder gauge --- and then imposing
a residual gauge condition on the solutions \cite{TW1}.

Grishchuk's result (\ref{EOM}) means that dynamical gravitons must
suffer from the same infrared problems as MMC scalars, not just in
de Sitter but for any spatially flat, FRW cosmology with $q \leq
+\frac12$. This can be seen by studying any one of the standard
2-point functions: the retarded Green's function, the Wightman
function or the Feynman propagator,
\begin{eqnarray}
\Bigl[ \mbox{}_{ij} G_{k\ell}\Bigr](x;x') & \equiv & -i \theta(\eta
\!-\! \eta') \Bigl[ \widetilde{h}_{ij}(\eta,\vec{x}) ,
\widetilde{h}_{k\ell}(\eta',\vec{x}') \Bigr] \; , \label{Green} \\
\Bigl[ \mbox{}_{ij} W_{k\ell}\Bigr](x;x') & \equiv & \Bigl\langle
\Omega \Bigl\vert \widetilde{h}_{ij}(\eta,\vec{x})
\widetilde{h}_{k\ell}(\eta',\vec{x}') \Bigr\vert \Omega \Bigr\rangle
\; , \label{Wightman} \\
i \Bigl[ \mbox{}_{ij} \Delta_{k\ell}\Bigr](x;x') & \equiv &
\Bigl\langle \Omega \Bigl\vert T\Bigl[
\widetilde{h}_{ij}(\eta,\vec{x})
\widetilde{h}_{k\ell}(\eta',\vec{x}') \Bigr] \Bigr\vert \Omega
\Bigr\rangle \; . \label{Feynman}
\end{eqnarray}
(Here $\vert \Omega \rangle$ stands for Bunch-Davies vacuum and $T$
denotes the time-ordering symbol.) Although the mode sum for the
retarded Green's function exists, its properties imply infrared
divergences for the Wightman function and for the propagator, which
is confirmed by direct examination of the mode sums for these
quantities.

Before proceeding it is useful to digress on the properties of the
three 2-point functions. Each of them obeys a simple equation when
acted upon by the scalar d'Alembertian,
\begin{eqnarray}
\sqrt{-g} \, \square \Bigl[ \mbox{}_{ij} G_{k\ell}\Bigr](x;x') & = & \Bigl[
\Pi_{i (k} \Pi_{\ell) j} - \frac{ \Pi_{ij} \Pi_{k\ell}}{d
\!-\! 2} \Bigr] \delta^d(x \!-\! x') \; , \\
\sqrt{-g} \, \square \Bigl[ \mbox{}_{ij} W_{k\ell}\Bigr](x;x')
& = & 0 \; , \\
\sqrt{-g} \, \square i \Bigl[ \mbox{}_{ij} \Delta_{k\ell}\Bigr](x;x') & = &
\Bigl[ \Pi_{i (k} \Pi_{\ell) j} - \frac{ \Pi_{ij} \Pi_{k\ell}}{d
\!-\! 2} \Bigr] i \delta^d(x \!-\! x') ; ,
\end{eqnarray}
where the transverse projection operator $\Pi_{ij}$ is defined by
Fourier transform just as in flat space,
\begin{equation}
\Pi_{ij} \equiv \delta_{ij} - \frac{ \partial_i
\partial_j}{\partial_k \partial_k} \; .
\end{equation}
To prevent indices (and the transverse-traceless projection
operator) from becoming an issue, we contract them across the
$x^{\mu}$ and ${x'}^{\mu}$ index groups,
\begin{equation}
G(x;x') \!\equiv\! \Bigl[ \mbox{}_{ij} G_{ij}\Bigr](x;x') \, , \, W(x;x')
\!\equiv\! \Bigl[ \mbox{}_{ij} W_{ij}\Bigr](x;x') \, , \, i\Delta(x;x')
\!\equiv\! i \Bigl[ \mbox{}_{ij} \Delta_{ij}\Bigr](x;x') \, .
\label{contract}
\end{equation}
One then sees that these contracted graviton 2-point functions obey
precisely the same equations as $\frac12 d (d-3)$ times the
analogous MMC scalar 2-point function, which are $\sqrt{-g} \, \square
W(x;x') = 0$ and,
\begin{equation}
i \sqrt{-g} \, \square G(x;x') = \frac12 d (d\!-\! 3) i \delta^d(x \!-\!
x') = \sqrt{-g} \, \square i\Delta(x;x') \; . \label{propeqn}
\end{equation}
Of course this is because each of the $\frac12 d (d-3)$ graviton
polarizations behaves like an independent MMC scalar.

Grishchuk's result poses a terrible obstacle for those who believe
that a de Sitter invariant state (as opposed to a possibly
non-normalizable wave function) exists for dynamical gravitons.
Although the unique solution for $G(x;x')$ follows from classical
general relativity and is de Sitter invariant, it cannot be
analytically continued to give either a de Sitter
invariant Wightman function or a de Sitter invariant propagator.
To see this, first compare expressions (\ref{Green}) and (\ref{Feynman})
to conclude,
\begin{equation}
G(x;x') = 2 \theta(\eta \!-\! \eta') {\rm Im}\Bigl[ i\Delta(x;x')\Bigr] \; .
\end{equation}
In $d=4$ dimensions the retarded Green's function is \cite{TW2},
\begin{equation}
\lim_{d \rightarrow 4} \frac{G(x;x')}{\frac12 d (d \!-\! 3)} =
-\frac{\theta(\eta \!-\! \eta')}{8 \pi \ell^2} \Biggl\{ \delta\Bigl(
z(x;x') \!-\! 1 \Bigr) + 2 \theta\Bigl(z(x;x') \!-\! 1 \Bigr)
\Biggr\} \; , \label{Gret}
\end{equation}
where $1- z(x;x') \equiv \frac14 a(\eta) a(\eta') \ell^{-2} (x - x')^{\mu}
(x- x')^{\nu} \eta_{\mu\nu}$ is a de Sitter invariant function of the
proper length from $x^{\mu}$ to ${x'}^{\mu}$. It is possible to get
both the delta function and the theta function from simple analytic
continuations,
\begin{equation}
\delta(z \!-\! 1) = -\frac1{\pi}
{\rm Im}\Bigl[ \frac1{1 \!-\! z \!+\! i \epsilon} \Bigr] \quad , \quad
\theta( z \!-\! 1) = \frac1{\pi} {\rm Im}\Bigl[ \ln(1 \!-\! z \!+\!
i\epsilon) \Bigr] \; .
\end{equation}
However, the resulting function does not obey the propagator equation
(\ref{propeqn}). The most general solution consistent with homogeneity
and isotropy takes the form \cite{TW2},
\begin{equation}
\lim_{d \rightarrow 4} \frac{i\Delta(x;x')}{\frac12 d (d \!-\! 3)}
= \frac1{16 \pi^2 \ell^2} \Biggl\{ \frac1{1 \!-\! z(x;x') \!+\! i \epsilon}
- 2 \ln\Biggl[ \frac{1 \!-\! z(x;x') \!+\! i \epsilon}{a(\eta) a(\eta')}
\Biggr] + {\rm Const.} \Biggr\} . \label{solution}
\end{equation}

The explicit dependence of (\ref{solution}) upon the scale factors is an
example of the secular behavior mentioned previously. This kind of secular
dependence has nothing special to do with de Sitter
and is a feature of all spatially flat, FRW cosmologies with
$q \leq +\frac12$ \cite{JMPW}. Note also that the de Sitter breaking part
of (\ref{solution}) drops out from the mixed second derivative
$\partial_{\mu} \partial_{\nu}' i\Delta(x;x')$, which is why the
expectation value of the MMC scalar stress tensor is de Sitter invariant,
even though the propagator is not. The same thing occurs for the tree order
Weyl-Weyl correlator and in no way implies that dynamical gravitons are
de Sitter invariant.

The argumentation just reviewed would seem decisive, but
mathematical physicists have for years ignored it because they were able
to derive manifestly de Sitter invariant expressions for the graviton
propagator by adding covariant gauge fixing terms to the Euclidean action
and then analytically continuing \cite{INVPROP}. It was recently shown
that both procedures are invalid, the first because it adds spurious zero
modes \cite{MTW1} and the second because it automatically subtracts power
law infrared divergences \cite{MTW2}. Further, an explicitly de Sitter
breaking result for the propagator has been derived in a covariant exact
gauge without subtracting infrared divergences \cite{MTW3}.

A sort of compromise has recently been suggested, based on some old
work by Higuchi \cite{oldwork}. The idea is that dynamical gravitons on de
Sitter might be physically de Sitter invariant, even though no manifestly
de Sitter invariant propagator for them can be found \cite{fallback}.
This proposal entails making sense of Bunch-Davies vacuum, which is the
unique possibility for a de Sitter invariant state for dynamical gravitons,
and that means somehow avoiding the infrared divergence. Recall that
the scale invariance of the graviton power spectrum, which
becomes exact in the de Sitter limit, implies infrared divergences in
the mode sums for both the Bunch-Davies Wightman function and for the
Bunch-Davies propagator \cite{KOW}.

It does not seem possible to argue that the scale invariance of the graviton
power spectrum is a gauge artifact. This is not only an observable quantity
but one which the Planck satellite is even now attempting to observe.
Nevertheless, Higuchi, Marolf and Morrison have recently devised an
ingenious argument that the infrared divergences of the mode sums for
$W(x;x')$ and $i\Delta(x;x')$ can be eliminated by making what seems to be
a gauge transformation which preserves the transverse-traceless-synchronous
conditions (\ref{TTS}) \cite{Don}. The purpose of this paper is to
critically examine that argument, which we review in section 2. In section
3 we demonstrate that the transformation changes both the propagator
equation and the canonical commutation relations. We also show that the same
procedure could be used to avoid the conclusion that no de Sitter invariant
state exists for the MMC scalar. A very similar gauge transformation even
seems to eliminate the infrared divergences of flat space QED. Our
discussion comprises section 4.

\section{The Transformation}

With one major exception, we have followed the notation of Higuchi,
Marolf and Morrison \cite{Don}, except for correcting a few typoes,
rearranging some factors and making the spacetime dependence
of the mode functions explicit. The major notational change is that
we use a tilde to denote conformally rescaled fields whereas they
employ it to represent their transformed fields. We use a prime to
indicate transformed fields.

The mode functions for a graviton with wave vector $\vec{k}$ and
spin $s$ take the form,
\begin{equation}
\gamma^s_{ij}(\eta,\vec{x};\vec{k}) = a^2(\eta) \times \frac{N
\sqrt{\ell}}{a^{\frac{d-1}2}} \, H^{(2)}_{\frac{d-1}2}(k\eta)
e^{i\vec{k} \cdot \vec{x}} \times \epsilon^s_{ij}(\vec{k}) \equiv
a^2(\eta) \widetilde{\gamma}^s_{ij}(\eta,\vec{x};\vec{k}) \; .
\label{modes}
\end{equation}
Here the normalization constant is,
\begin{equation}
N \sqrt{\ell} \equiv \overline{N} = \sqrt{\frac{\ell \pi}{4
(2\pi)^{d-1}}} \; , \label{norm}
\end{equation}
and the transverse-traceless polarization tensors are the same as
those of flat space,
\begin{equation}
k_i \epsilon^s_{ij}(\vec{k}) = 0 = \epsilon^s_{ii}(\vec{k}) \qquad ,
\qquad \epsilon^r_{ij}(\vec{k}) \epsilon^{s*}_{ij}(\vec{k}) =
\delta^{rs} \; . \label{polar}
\end{equation}
It is worth noting that the central term in expression (\ref{modes})
is the mode function for a massless, minimally coupled scalar with
wave vector $\vec{k}$,
\begin{equation}
u(\eta,\vec{x};\vec{k}) = \frac{\overline{N}}{a^{\frac{d-1}2}} \,
H^{(2)}_{\frac{d-1}2}(k\eta) e^{i\vec{k} \cdot \vec{x}} = \frac{i
\overline{N}}{\pi} \Gamma\Bigl(\frac{d\!-\!1}2\Bigr)
\Bigl(\frac2{\ell k}\Bigr)^{\frac{d-1}2} \Bigl\{ 1 + O(k^2 \eta^2)
\Bigr\} e^{i \vec{k} \cdot \vec{x}} \; . \label{MMCS}
\end{equation}

Mode functions are $\comp$-number solutions of the linearized field
equations. The operator which annihilates a graviton with wave
number $\vec{k}$ and spin $s$ is $a_s(\vec{k})$. The nonzero part of
the commutation algebra is,
\begin{equation}
\Bigl[ a_r(\vec{k}) , a_s^{\dagger}(\vec{p}) \Bigr] = \delta^{rs}
\delta^{d-1}(\vec{k} \!-\! \vec{p}) \; . \label{CandA}
\end{equation}
The mode sum for the transverse-traceless-synchronous graviton field
operator is,
\begin{equation}
h_{ij}(\eta,\vec{x}) = a^2(\eta) \int \!\! d^{d-1}\!k \sum_s
\Biggl\{ \widetilde{\gamma}^s_{ij}(\eta,\vec{x};\vec{k})
a_s(\vec{k}) + \widetilde{\gamma}^{s*}_{ij}(\eta,\vec{x};\vec{k})
a_s^{\dagger}(\vec{k}) \Biggr\} \; . \qquad \label{modesum}
\end{equation}
Note that raising one index with the full metric gives the
conformally rescaled graviton field, whose indices are raised and
lowered with $\delta_{ij}$,
\begin{equation}
h^i_{~j}(\eta,\vec{x}) \equiv g^{ik} h_{kj}(\eta,\vec{x}) =
\widetilde{h}_{ij}(\eta,\vec{x}) \label{conf}
\end{equation}

The full equal-time commutation algebra of the fields is,
\begin{eqnarray}
\Bigl[ \widetilde{h}_{ij}(\eta,\vec{x}) ,
\widetilde{h}_{k\ell}(\eta,\vec{y}) \Bigr] & = & 0
= \Bigl[ \partial_0 \widetilde{h}_{ij}(\eta,\vec{x}) ,
\partial_0 \widetilde{h}_{k\ell}(\eta,\vec{y}) \Bigr] \; ,
\label{com12} \\
\Bigl[ \widetilde{h}_{ij}(\eta,\vec{x}) , \partial_0
\widetilde{h}_{k\ell}(\eta,\vec{y}) \Bigr] & = &
\Bigl[ \Pi_{i (k} \Pi_{\ell) j} - \frac{ \Pi_{ij}
\Pi_{k\ell}}{d \!-\! 2} \Bigr] \frac{i \delta^{d-1}( \vec{x} \!-\!
\vec{y})}{a^{d-2}} \; . \label{com3}
\end{eqnarray}
Of course fields are the primary objects of quantum field
theory so one really starts with the field equation (\ref{EOM})
for $\widetilde{h}_{ij}(\eta,\vec{x})$. That implies the form
(\ref{modesum}) with relations (\ref{modes}-\ref{polar}). The
Fourier coefficients $a_s(\vec{k})$ and $a^{\dagger}_s(\vec{k})$
of each mode function are properly combinations of the initial
values of the field and its first time derivative. The equal-time
commutation relations (\ref{com12}-\ref{com3}) of these fields
imply (\ref{CandA}).

The unique de Sitter invariant wave function for this field is
Bunch-Davies vacuum defined by the condition,
\begin{equation}
a_s(\vec{k}) \Bigl\vert \Omega\Bigr\rangle = 0 \quad \forall \;
\vec{k} \; {\rm and} \; s \; .
\end{equation}
If we incorrectly assume $\langle \Omega \vert \Omega \rangle = 1$
it is simple to show that the contracted graviton Wightman function
defined by expressions (\ref{Wightman}) and (\ref{contract}) is
just $\frac12 d (d-3)$ copies of the Wightman
function for a massless, minimally coupled scalar,
\begin{eqnarray}
W(x;x') & = & \int \!\! d^{d-1}\!k \sum_s
\widetilde{\gamma}^s_{ij}(\eta,\vec{x};\vec{k})
\widetilde{\gamma}^{s*}_{ij}(\eta',\vec{x}' ;\vec{k}) \; , \qquad
\label{W2grav} \\
& = & \frac12 d (d \!-\! 3) \int \!\! d^{d-1}\!k \,
u(\eta,\vec{x};\vec{k}) u^*(\eta',\vec{x}' ; \vec{k}) \; . \qquad
\label{W2scal}
\end{eqnarray}
Owing to the small $k$ behavior of the scalar mode functions
(\ref{MMCS}), expression (\ref{W2scal}) fails to converge in the
infrared. The usual interpretation is that the single assumption
from which these relations were derived --- $\langle \Omega \vert
\Omega \rangle = 1$ --- is false and, like the massless, minimally
coupled scalar, there are no normalizable, de Sitter invariant
states for the graviton.

Higuchi, Marolf and Morrison have argued that this conclusion is
wrong \cite{Don}. They assert that the infrared divergence of
(\ref{W2grav}-\ref{W2scal}) is pure gauge and can be eliminated by
what seems to be a linearized gauge transformation. The
transformation parameters are $\xi_0(\eta,\vec{x}) = 0$ and,
\begin{equation}
\xi_i(\eta,\vec{x}) = - a^2(\eta) \int \!\! d^{d-1}\!k \, e^{-\rho
k} \sum_s \Biggl\{ \widetilde{\gamma}^s_{ij}(0,\vec{0};\vec{k})
a_s(\vec{k}) + \widetilde{\gamma}^{s*}_{ij}(0,\vec{0};\vec{k})
a^{\dagger}_s(\vec{k}) \Biggr\} x^i \; , \label{oldxi}
\end{equation}
where $\rho$ is some positive constant. The transformed graviton
field is,
\begin{equation}
h_{\mu\nu}'(\eta,\vec{x}) \equiv h_{\mu\nu}(\eta,\vec{x}) +
\nabla_{(\mu} \xi_{\nu)}(\eta,\vec{x}) \; . \label{transform}
\end{equation}
The new fields also obey the transverse-traceless-synchronous
conditions (\ref{TTS}). The mode sums for their nonzero
components are,
\begin{eqnarray}
\lefteqn{\widetilde{h}_{ij}'(\eta,\vec{x}) = \int \!\! d^{d-1}\!k \sum_s
\Biggl\{ \Bigl[\widetilde{\gamma}^s_{ij}(\eta,\vec{x};\vec{k}) -
e^{-\rho k} \widetilde{\gamma}^s_{ij}(0,\vec{0};\vec{k}) \Bigr]
a_s(\vec{k}) } \nonumber \\
& & \hspace{4.8cm} +
\Bigl[\widetilde{\gamma}^{s*}_{ij}(\eta,\vec{x};\vec{k}) - e^{-\rho
k} \widetilde{\gamma}^{s*}_{ij}(0,\vec{0};\vec{k}) \Bigr]
a_s^{\dagger}(\vec{k}) \Biggr\} \; . \qquad \label{newsum}
\end{eqnarray}
It is immediately apparent that the new mode functions are better
behaved near $k = 0$, so that the mode sum for the transformed Wightman
function converges,
\begin{eqnarray}
\lefteqn{W'(x;x') = \frac12 d (d \!-\! 3) } \nonumber \\
& & \hspace{.5cm} \times \int \!\! d^{d-1}\!k \,
\Bigl[u(\eta,\vec{x};\vec{k}) \!-\! e^{-\rho k} u(0,\vec{0};\vec{k})
\Bigr] \Bigl[ u^*(\eta',\vec{x}' ; \vec{k}) \!-\! e^{-\rho k}
u(0,\vec{0}; \vec{k}) \Bigr] \; . \qquad \label{newW2}
\end{eqnarray}
Because the change from $h_{ij}(\eta,\vec{x})$ to
$h_{ij}'(\eta,\vec{x})$ was a gauge transformation, Higuchi, Marolf
and Morrison conclude that no observable quantity is affected.
Because Bunch-Davies vacuum now exists, they conclude that the
graviton vacuum is de Sitter invariant.

Higuchi, Marolf and Morrison were aware of the dangers associated
with employing a transformation parameter (\ref{oldxi}) which grows
at spatial infinity. They therefore offered a local version of their
argument in which the transformation parameter is made to vanish at
spatial infinity \cite{Don}. The method consists of multiplying the
old transformation parameter by a ``tophat'' function of space and
time,
\begin{equation}
\xi^{\rm new}_i(\eta,\vec{x}) \equiv \xi^{\rm old}_{i}(\eta,\vec{x})
\times T(\eta,\vec{x}) \; . \label{newxi}
\end{equation}
The tophat function $T(\eta,\vec{x})$ is chosen to be unity within
some arbitrarily large but finite region of space, and to vanish
outside some larger but still finite region. Then the transformed
graviton field agrees with (\ref{newsum}) inside the smaller region,
and the corresponding Wightman function is infrared finite
within this region. Although the Wightman function diverges outside
the larger region, they assert that this is unobservable because the
smaller region can be made arbitrarily large. They even consider
allowing the smaller region to encompass the past light-cone of some
finite point back to $\eta \rightarrow -\infty$.

\section{Problems with the Transformation}

Let us first observe that the Wightman function of the
transformed field has a simple interpretation in terms of the
original Wightman function,
\begin{eqnarray}
\lefteqn{W'(x;x') \!=\! W(x;x') \!-\! \int \!\! d^{d-1}\!z \,
F(\vec{z},\rho) W(x;0,\vec{z}) \!-\! \int \!\! d^{d-1}\!y \,
F(\vec{y},\rho) W(0,\vec{y};x') } \nonumber \\
& & \hspace{3.9cm} + \int \!\! d^{d-1}\!y \, F(\vec{y},\rho) \int
\!\! d^{d-1}\!z \, F(\vec{z},\rho) W(0,\vec{y};0,\vec{z}) \; .
\qquad \label{4terms}
\end{eqnarray}
Here the real function $F(\vec{y},\rho)$ is defined by Fourier
transforming $e^{-\rho k}$,
\begin{equation}
F(\vec{y},\rho) \equiv \int \!\! \frac{d^{d-1}\!k}{(2 \pi)^{d-1}} \,
e^{-i \vec{k} \cdot \vec{y}} e^{-\rho k} \; .
\end{equation}
It is apparent that the transformed field is a sort of dipole with
anti-sources at $\eta = 0$ (which is the infinite future) to balance
its sources at $x^{\mu}$ and ${x'}^{\mu}$. That is the origin of its
better infrared behavior.

{\it The problem with adding anti-sources is that it makes the
transformed propagator obey a different equation from the original
one even though no gauge condition has been changed}. There are two
plausible ways of defining what is meant by the transformed propagator,
and neither of these definitions obeys (\ref{propeqn}). The simplest
definition is just to extend (\ref{4terms}),
\begin{eqnarray}
\lefteqn{i\Delta_1'(x;x') \!\equiv\! i\Delta(x;x') \!-\! \int \!\!
d^{d-1}\!z \, F(\vec{z},\rho) i\Delta(x;0,\vec{z}) \!-\!
\int \!\! d^{d-1}\!y \, F(\vec{y},\rho) i\Delta(0,\vec{y};x') }
\nonumber \\
& & \hspace{3.9cm} + \int \!\! d^{d-1}\!y \, F(\vec{y},\rho) \int
\!\! d^{d-1}\!z \, F(\vec{z},\rho) i\Delta(0,\vec{y};0,\vec{z}) \; .
\qquad \label{def1}
\end{eqnarray}
The equation obeyed by this propagator is,
\begin{equation}
\sqrt{-g} \, \square i\Delta_1'(x;x') = \frac12 d (d \!-\! 3)
\Bigl\{ i \delta^d(x \!-\! x') - i\delta(\eta) F(\vec{x},\rho)
\Bigr\} \; . \label{bad1}
\end{equation}

The more orthodox definition for the transformed propagator is,
\begin{equation}
i\Delta_2'(x;x') \equiv \theta(\eta \!-\! \eta') W'(x;x') +
\theta(\eta' \!-\! \eta) W'(x';x) \; . \label{def2}
\end{equation}
The equation it obeys is revealing,
\begin{eqnarray}
\lefteqn{ \sqrt{-g} \, \square i\Delta_2'(x;x') = [a(\eta')]^{d-2}
\Biggl\{ \delta(\eta \!-\! \eta') \Bigl[ \partial_0
W'(\eta',\vec{x}';\eta,\vec{x}) \!-\! \partial_0
W'(\eta,\vec{x} ;\eta',\vec{x}') \Bigr] } \nonumber \\
& & \hspace{3.5cm} + \delta'(\eta \!-\! \eta') \Bigl[
W'(\eta',\vec{x}' ;\eta',\vec{x}) \!-\! W'(\eta',\vec{x} ;
\eta',\vec{x}') \Bigr] \Biggr\} \; , \qquad \label{1stline} \\
& & = \frac12 d (d\!-\!3) i\delta^d(x \!-\! x') + i [a(\eta')]^{d-2}
\delta'(\eta \!-\! \eta') \nonumber \\
& & \hspace{3.5cm} \times \int \!\! d^{d-1}\!y \, F(\vec{y},\rho)
\Bigl[ G(0,\vec{y};\eta',\vec{x}) \!-\! G(0,\vec{y};\eta',\vec{x}')
\Bigr] \; . \qquad \label{2ndline}
\end{eqnarray}
Higuchi, Marolf and Morrison checked that the equal-time commutator
of the transformed field and its first time derivative on the first
line of (\ref{1stline}) is unchanged from (\ref{com3}) \cite{Don},
\begin{eqnarray}
\lefteqn{[a(\eta')]^{d-2} \delta(\eta \!-\! \eta') \Bigl[ \partial_0
W'(\eta',\vec{x}';\eta,\vec{x}) \!-\! \partial_0
W'(\eta,\vec{x};\eta',\vec{x}') \Bigr] } \nonumber \\
& & \hspace{1.5cm} = [a(\eta')]^{d-2} \delta(\eta \!-\! \eta')
\Bigl[ \partial_0 W(\eta',\vec{x}';\eta,\vec{x}) \!-\! \partial_0
W(\eta,\vec{x};\eta',\vec{x}') \Bigr] \; , \qquad \\
& = & \frac12 d (d\!-\!3) i \delta^{d-1}(\vec{x} \!-\!
\vec{x'}) \; . \qquad
\end{eqnarray}
The extra term in (\ref{2ndline}) comes from the equal-time commutator
of the undifferentiated fields. Higuchi, Marolf and Morrison seem to
have assumed that this vanishes, just as its does for the original
fields (\ref{com12}). That would be a safe assumption for a local
transformation, but it is false for this nonlocal transformation. {\it
The source of the extra term in (\ref{2ndline}) is a change in the
canonical commutation relations (\ref{com12}).}

Additional evidence is provided by the retarded Green's function
of the transformed field. If this is defined from the transformed
Wightman function in the way one would normally do, assuming
the canonical commutation relations (\ref{com12}-\ref{com3}), one
finds that it changes,
\begin{eqnarray}
G'(x;x') & \equiv & -i\theta(\eta \!-\! \eta') \Bigl[
W'(x;x') \!-\! W'(x';x) \Bigr] \; , \label{Gdef} \\
& = & G(x;x') + \theta(\eta \!-\! \eta') \int \!\! d^{d-1}\!y \,
F(\vec{y},\rho) \Bigl[ G(0,\vec{y};x) \!-\! G(0,\vec{y};x')\Bigr]
\; . \qquad
\end{eqnarray}
But the retarded Green's function follows from the unchanged field
equations of classical general relativity and cannot change. Of course
it does not change; the problem is that its relation to the Wightman
function (\ref{Gdef}) assumes the canonical commutations relations
(\ref{com12}-\ref{com3}), which are not correct for the transformed
field.

The reason the canonical commutation relations changed is obvious in
retrospect. The transverse-traceless-synchronous conditions (\ref{TTS})
{\it already} fix the gauge completely; there is no additional local
gauge freedom. One can see that from the fact that the field equation
(\ref{EOM}) and the canonical commutation relations (\ref{com12}-\ref{com3})
give a unique solution (\ref{modesum}). So any transformation which
alters the mode sum without disturbing the transverse-traceless-synchronous
condition (\ref{TTS}) can only do so by changing the canonical commutation
relations (\ref{com12}-\ref{com3}).

Transformation (\ref{transform}) shifts the spacetime constant part of the
field. This would not matter if the shift were by a $\comp$-number but it
matters a great deal when the shift is by a quantum operator because that
operator possesses correlations with other operators whose
spacetime dependence it can acquire. What (\ref{transform}) does is to load
up the constant part of the field --- which should have been an independent
degree of freedom --- with a superposition of creation and annihilation
operators of all different wave numbers and spins. This becomes obvious if
we rewrite the transformation parameter (\ref{oldxi}) in terms of the fields,
\begin{eqnarray}
\xi_i(\eta,\vec{x}) & = & -a^2(\eta) \int \!\! d^{d-1}\!y \, F(\vec{y},\rho)
\widetilde{h}_{ij}(0,\vec{y}) x^j \; , \qquad \label{charade} \\
\widetilde{h}_{ij}'(\eta,\vec{x}) & = & \widetilde{h}_{ij}(\eta,\vec{x}) -
\int \!\! d^{d-1}\!y F(\vec{y},\rho) \widetilde{h}_{ij}(0,\vec{y}) \; . \qquad
\end{eqnarray}
From these expressions we see that violations of the propagator equation
(\ref{propeqn}) and of the canonical commutation relations (\ref{com12})
occur in both the global version of the transformation (\ref{oldxi}) and
also the localized version (\ref{newxi}). This is significant because the
localized version does change the gauge conditions in the transition region
outside the smaller region and inside the larger region.

The altered commutation relations mean that Higuchi, Marolf and Morrison
have potentially changed the physics of dynamical gravitons, rather than
discovering a hitherto hidden property of the existing physics. The
conclusion for the original theory is therefore unchanged: there is no de
Sitter invariant state for dynamical gravitons. However, the willingness
of mathematical physicists to resort to such ingenious expedients to
salvage some form of de Sitter invariance for gravitons suggests that one
should examine the viability of changing canonical commutation as has been
done.

We can see three obvious problems. The first is that the new modes seem to
modify the tensor power spectrum,\footnote{The unfamiliar factor at the
beginning of expression (\ref{Pspecdef}) arises from the way Higuchi,
Marolf and Morrison define Fourier transforms and normalize
the graviton field.}
\begin{eqnarray}
\Delta^2_h(k) & \equiv & \Bigl[ 32\pi G (2\pi)^3 \Bigr]
\frac{k^3}{2 \pi^2} \!\! \int \!\! d^3x \, e^{-i \vec{k} \cdot \vec{x}}
\Bigl\langle \Omega \Bigl\vert \widetilde{h}_{ij}'(0,\vec{x})
\widetilde{h}_{ij}'(0,\vec{0}) \Bigr\vert \Omega \Bigr\rangle \; ,
\qquad \label{Pspecdef} \\
& = & \Bigl[32\pi G (2\pi)^3\Bigr] \frac{k^3}{2 \pi^2} \sum_s
\widetilde{\gamma}^s_{ij}(0,\vec{0};\vec{k})
\widetilde{\gamma}^{s*}_{ij}(0,\vec{0};\vec{k}) \times \Bigl[1 \!-\!
e^{-\rho k}\Bigr] \; , \qquad \\
& = & \frac{16 G}{\pi \ell^2} \times \Bigl[1 \!-\!
e^{-\rho k}\Bigr] \; . \qquad \label{Pspectrum}
\end{eqnarray}
The factor of $[1 - e^{-\rho k}]$ multiplying the usual de Sitter result in
expression (\ref{Pspectrum}) is more evidence that physics has been changed.
We are not told the numerical value of $\rho$ relative to observable
cosmological wave numbers $k$ but, if it is any natural scale of fundamental
theory --- from the Planck length to the Compton wavelength of a neutrino
--- then one has $[1 - e^{-\rho k}] \approx \rho k$ for observable
perturbations. That would give the graviton power spectrum a massive blue
tilt, corresponding to a spectral index of $n_T = +1$. The usual prediction
is $n_T$ negative and much less than one \cite{perts}; for single scalar
inflation the data implies $-0.028 < n_T < 0$ at 95\% confidence
\cite{KOW,WMAP}.

The second problem is that changing canonical quantization could
be done as well for the MMC scalar. Indeed, with the graviton indices
contracted across index groups as we have been doing, the graviton
relations are just $\frac12 d (d \!-\!3)$ copies of the MMC scalar
relations. Making the same change in the scalar commutations relations
would produce a propagator whose coincidence limit no longer agrees
with the classic result of Vilenkin and Ford, Linde and Starobinsky
\cite{VFLS}, or with the results that follow from it when interactions are
present \cite{phi4,Yukawa,SQED}. These problems are confined to the
realm of pure theory, whereas the ultimate justification of any physical
model should be how it explains the data. Here the decisive fact seems to
be that the scalar power spectrum would not be even approximately scale
invariant as with the tensor power spectrum), in conflict with
observation \cite{WMAP}.

The final problem is that the very same type of gauge
transformation changes the infrared behavior of the photon
propagator in flat space QED. Consider free dynamical
photons in Coulomb-synchronous gauge,
\begin{equation}
A_0(t,\vec{x}) = 0 \qquad , \qquad \partial_i A_i(t,\vec{x}) = 0 \; .
\label{CS}
\end{equation}
In $d=4$ spacetime dimensions the mode sum takes the form,
\begin{equation}
A_i(t,\vec{x}) = \int \!\! d^3k \sum_s \Biggl\{ \gamma^s_i(t,\vec{x};\vec{k})
a_s(\vec{k}) + \gamma^{s*}_i(t,\vec{x};\vec{k}) a^{\dagger}_s(\vec{k})
\Biggr\} \; , \label{photons}
\end{equation}
where the mode functions are,
\begin{equation}
\gamma^s_i(t,\vec{x};\vec{k}) = \frac1{\sqrt{(2\pi)^3 2 k}} \,
e^{-ik t + i \vec{k} \cdot \vec{x}} \epsilon^s_i(\vec{k}) \; .
\end{equation}
The polarization vectors $\epsilon^s_i(\vec{k})$ are the usual ones,
and the creation and annihilation operators obey the same relations
(\ref{CandA}) as those for dynamical gravitons. In direct analogy to
(\ref{charade}) we make a gauge transformation with parameter,
\begin{equation}
\theta(t,\vec{x}) = \int \!\! d^3y \, F(\vec{y},\rho) A_i(0,\vec{y}) x^i
\end{equation}
Just as was the case with (\ref{oldxi}), the transformed vector potential
continues to obey the Coulomb-synchronous conditions (\ref{CS}),
\begin{equation}
A_i'(t,\vec{x}) \equiv A_i(t,\vec{x}) - \partial_i \theta(t,\vec{x})
= A_i(t,\vec{x}) - \int \!\! d^3y \, F(\vec{y},\rho) A_i(0,\vec{y}) \; .
\end{equation}

As was the case with gravitons, the new photon mode sum is better behaved
in the infrared,
\begin{eqnarray}
\lefteqn{ A_i'(t,\vec{x}) = \int \!\! d^3k \sum_s \Biggl\{ \Bigl[
\gamma^s_i(t,\vec{x};\vec{k}) \!-\! e^{-\rho k} \gamma^s_i(0,\vec{0};\vec{k})
\Bigr] a_s(\vec{k}) } \nonumber \\
& & \hspace{5cm} + \Bigl[ \gamma^{s*}_i(t,\vec{x};\vec{k}) \!-\!
e^{-\rho k} \gamma^{s*}_i(0,\vec{0};\vec{k}) \Bigr] a_s^{\dagger}(\vec{k})
\Bigr] \Biggr\} . \qquad
\end{eqnarray}
Indeed, it seems to be too well behaved. The old mode sum involved an
integrable $1/k$ divergence which allows the position-space propagator
to be computed but leads to infrared divergences in loop corrections to
exclusive scattering amplitudes. Those divergences are not some sort of
gauge artifact, they arise from our inability to measure exclusive
processes. When the problem is corrected by shifting to inclusive
amplitudes the infrared behavior of the photon propagator leads to
an observable dependence on the detector sensitivity that agrees with
experiment \cite{BNSW}. None of this seems to be true for the new mode
sum because it fails to diverge at $k =0$.

One might object that QED scattering amplitudes are
gauge invariant, so they cannot be changed by making a gauge
transformation. This argument is false because the gauge transformation
was not made on the full, interacting theory. Indeed, it {\it cannot}
be made on the full theory because the transformation exploited special
properties of the free theory such as (\ref{CS}-\ref{photons}) that are
no longer true in the presence of interactions. What is being done
instead is to change the canonical commutation relations, and that
generally does change physics. The problem is that it seems to change
things we have already measured.

\section{Discussion}

As we have seen, what appears to be a linearized gauge transformation
is actually a way of altering the canonical commutation relations
(\ref{com12}-\ref{com3}) of dynamical gravitons in
transverse-traceless-synchronous gauge. This is because adding a
$\comp$-number constant is not the same as adding a constant operator.
The correlations of the latter with other operators can permit it to
inherit their spacetime dependence.

One consequence is that the theory of dynamical gravitons is being
changed, rather than that some hidden property of the usual theory
is being uncovered. The conclusion for the original theory is unambiguous:
the Bunch-Davies wave function is not normalizable for any spatially flat
cosmology with deceleration parameter $q \leq +\frac12$. Because de Sitter
(with $q = -1$) is just one of a wide range of cosmologies for which
dynamical gravitons possess infrared problems, one should not opine that
the resolution of its infrared problem would be ever so much clearer on
the full de Sitter manifold. No such appeal is possible for cosmologies
with $-1 < q \leq +\frac12$, so any argument about the infrared problem
being a gauge artifact must make sense for spatially flat, FRW cosmologies
with $q \leq +\frac12$.

Is the new theory an object of interest? Many strange things happen
when theories are not quantized canonically. Not all of these things are
provably wrong but this particular procedure seems to be. It could be
used to avoid the infrared divergence of the massless, minimally coupled
scalar and seems to change both the tensor and scalar power spectra.
For flat space QED the same sort of transformation seems to eliminate
the infrared divergences of the exclusive scattering amplitudes. These
things have measurable consequences and cannot be altered. The safest
conclusion at this stage is that extraordinary claims require extraordinary
proofs. Someone asserting that the new commutation relations give an
acceptable theory (which no one has) should supply this proof, and that
must include more than observing that the new relations were obtained by
making what seems to be a gauge transformation.

Our comments apply to both the global version (\ref{oldxi}) and the
localized version (\ref{newxi}) of the transformation. However, it should
be noted that the localized version of the transformation is not relevant
to the discussion of de Sitter invariance. By definition, de Sitter
invariance is a global property of the full state, not a local one, so
no procedure which leaves the propagator divergent outside some arbitrary
but large region could be invoked to establish it.

If it could be done, establishing the de Sitter invariance of dynamical
gravitons would seriously alter how individual particles propagate through
the vast ensemble of infrared gravitons which is created by inflation --- at
least in the usual physics. This problem has been studied at one loop order
for massless fermions \cite{MW1} and for MMC scalars \cite{KW}. The results
are that MMC scalars suffer no significant correction whereas massless
fermions experience a secular enhancement of their field strengths
that eventually becomes nonperturbatively strong. The difference between
the two particles seems to be spin; MMC scalars only interact with inflationary
gravitons through their kinetic energies, which rapidly redshift to zero.
Fermions experience an additional interaction due to spin, and it is precisely
the diagrams involving the spin connection which are responsible for the one
loop effect \cite{MW2}. That suggests there should be a similarly strong effect
for gravitons and photons. Neither computation has been made yet but the
presence of infrared logarithms (which are not present in the altered theory of
Higuchi, Marolf and Morrison \cite{Don}) in the one loop graviton self-energy
argues that they also experience a strong effect \cite{TW3}.

Although we do not believe that even free dynamical gravitons are de
Sitter invariant, it should be noted that their propagator does not give
the full gravitational response to a source, nor even the most important
part of this response. In a covariant gauge such as de Donder \cite{MTW3}
the full propagator includes the effects of constraints as well as of
dynamical gravitons. Of the classic tests of general relativity, all
but the spin-down of the binary pulsar have to do with the
constrained fields, not with the dynamical ones. This comment is especially
significant for the quantum gravitational back-reaction on inflation because
the physical mechanism proposed for it \cite{TW4} involves the gravitational
response to the source provided by the continual inflationary production
of dynamical gravitons. The stress-energy of these gravitons does not break
de Sitter invariance at lowest order \cite{TW5}, any more than the
stress-energy of the MMC scalar does. The lowest order secular back-reaction
derives from the way the constrained fields respond to more and more
inflationary gravitons appearing inside the past light-cone.

It should also be noted that even the manifest de Sitter invariance of the
full graviton propagator (which no one seems to be maintaining any longer)
would not preclude secular back-reaction from loop corrections.
These loop corrections involve integrations of products of
propagators and vertex operators. A de Sitter invariant propagator just
means that the {\it integrands} are de Sitter invariant; the result of
performing the {\it integrals} can still break de Sitter invariance
\cite{RPW}. A classic example is the function $\theta(-\ell^2(x;x'))$,
where $\ell(x;x')$ is the invariant distance from $x^{\mu}$ to ${x'}^{\mu}$.
This function is certainly invariant: it is one inside the light-cone and
zero outside. Yet its integral over ${x'}^{\mu}$ to the past of $x^{\mu}$,
back to the initial value surface, gives the invariant volume of the past
light-cone. That quantity grows as $x^{\mu}$ is evolved to the
future. This example is not specious. Precisely this theta
function appears as the ``tail term'' of the retarded graviton
propagator (\ref{Gret}).

\centerline{\bf Acknowledgements}

We are grateful to D. Marolf for his courtesy in bringing an early
version of his paper to our attention, and for patiently answering
our questions and comments over the course of many e-mail exchanges.
We also thank S. Deser , D. Marolf and I. Morrison for reading a
preliminary version of this paper and providing valuable comments.
This work was partially supported by NWO Veni Project \# 680-47-406,
by European Union Grant FP-7-REGPOT-2008-1-CreteHEPCosmo-228644, by
NSF grant PHY-0855021, and by the Institute for Fundamental Theory
at the University of Florida.

\end{document}